\documentclass[pmlr]{jmlr}

\usepackage{amsmath,amssymb,graphicx,url}
\jmlrvolume{329}
\jmlryear{2026}
\jmlrworkshop{Conformal and Probabilistic Prediction with Applications}
\jmlrproceedings{PMLR}{Proceedings of Machine Learning Research}

\usepackage{longtable}
\usepackage{booktabs}
\usepackage[load-configurations=version-1]{siunitx} 

\usepackage[capitalise]{cleveref}
\crefname{figure}{Figure}{Figures} 
\Crefname{figure}{Figure}{Figures}

\usepackage{caption}
\usepackage[printonlyused]{acronym}
\newacro{GW}{gravitational wave}
\newacro{CP}{conformal prediction}
\newacro{SNR}{signal-to-noise ratio}
\newacro{FAR}{false alarm rate}
\newacro{IFAR}{inverse false alarm rate}
\newacro{TP}{true positive}
\newacro{FP}{false positive}
\newacro{FN}{false negative}
\newacro{TN}{true negative}
\newacro{TPR}{true positive rate}
\newacro{FPR}{false positive rate}
\newacro{ML}{machine learning}
\newacro{CNN}{convolutional neural network}
\newacro{ROC}{receiver operating characteristic}
\newacro{AUC}{area under the curve}
\newacro{MLP}{multi-layer perceptron}
\newacro{NN}{neural network}
\newacro{KNN}{K-nearest neighbours}
\newacro{LR}{logistic regression}
\newacro{BNS}{binary neutron star}
\newacro{BBH}{binary black hole}
\newacro{NSBH}{neutron star black hole}
\newacro{CBC}{compact binary coalescence}
\newacro{LVK}{LIGO-Virgo-KAGRA}
\newacro{SHAP}{SHapley Additive exPlanations}
\newacro{MDC}{mock data challenge}
\newacro{LLPIC}{low-latency production integration challenge}
\newacro{EMD}{earth mover's distance}

\newcommand{\maxifar}{maximum-$\log_{10}$IFAR\xspace}

\newcommand{\sklearn}{\texttt{scikit-learn}\xspace}
\newcommand{\pycbc}{\texttt{PyCBC}\xspace}
\newcommand{\gstlal}{\texttt{GstLAL}\xspace}
\newcommand{\mbta}{\texttt{MBTA}\xspace}
\newcommand{\cwb}{\texttt{CWB}\xspace}

\title[Gravitational Wave Detection with Weighted Conformal Prediction]{Improving the Sensitivity of Gravitational Wave Detection with Weighted Conformal Prediction}

\author{\Name{Ann-Kristin Malz} \Email{ann-kristin.malz@ligo.org}\\
 \addr Department of Physics, Royal Holloway, University of London
 \AND
 \Name{Gregory Ashton} \Email{gregory.ashton@ligo.org}\\
 \addr School of Mathematical Sciences, University of Southampton
 \AND
 \Name{Nicolo Colombo} \Email{nicolo.colombo@rhul.ac.uk}\\
 \addr Department of Computer Science, Royal Holloway University of London
 }

\editor{Ernst Ahlberg, Ulf Johansson, Henrik Boström, Alberto Carlevaro, Johan Hallberg Szabadváry and Lars Carlsson}

\begin{document}
\maketitle

\begin{abstract}
In the last decade, kilometre-scale interferometric gravitational-wave detectors have observed hundreds of compact binary mergers, the majority of which are binary black holes.
However, the data are noise-dominated, and multiple independent search algorithms (pipelines) are used to enhance sensitivity and improve robustness. Rather than the standard approach of selecting the most significant pipeline output, we combine the outputs from all pipelines using a \acl{CP}-based framework to provide statistically rigorous confidence estimates for candidate events. While combining pipelines improves sensitivity and ranking robustness, it requires a principled statistical framework that remains valid as data properties evolve across observing runs. A key challenge is distribution shifts between simulated datasets used for training and calibration and the real, unlabelled, observations used for testing, which can invalidate coverage guarantees and bias confidence estimates. In this work, we address this challenge by incorporating likelihood-ratio reweighting into our \acl{CP} framework to account for covariate shift. Using mock datasets containing simulated signals, we demonstrate that weighted \acl{CP} restores well-calibrated coverage under covariate shift and increases the confidence of events near the detection threshold, recovering true signals that would otherwise be missed. 
\end{abstract}


\acresetall  

\section{Introduction} \label{sec:intro}
\Acp{GW} are ripples in spacetime generated by accelerating massive objects, such as merging black holes or neutron stars. Their detection enables the study of astrophysical systems that are otherwise difficult or impossible to observe, offering a unique window into some of the most energetic events in the Universe and allowing us to test physical theories under extreme conditions. In practice, these signals are extremely weak and are embedded in noisy time-series data collected by ground-based interferometric detectors \citep{weiss1972electromagnetically}.

The field of \ac{GW} astronomy has grown rapidly since the first detection in 2015 \citep{LIGOScientific:2016aoc}, and with the latest catalogue release, GWTC-5.0~\citep{GWTC5}, the cumulative catalogue of events now contains almost 400 signals from \acl{CBC} sources observed by the \ac{LVK} detector network~\citep{LIGO, Virgo, KAGRA}. The majority of these detections correspond to \acl{BBH} mergers, with a small number originating from \acl{NSBH} and \acl{BNS} systems. 

\ac{GW}s are detected by the \ac{LVK} detectors, which are kilometre-scale laser interferometers that measure differential arm length variations induced by passing gravitational waves. For a review of \ac{GW} data analysis, see \citet{LIGOScientific:2019hgc}. 
The data are continuously analysed using dedicated search algorithms (pipelines) that scan for candidate events. Some pipelines are based on matched filtering, searching for known signal morphologies by correlating the data against precomputed waveform templates, while others use unmodelled or coherence-based approaches that make no assumptions about the signal shape \citep{LIGOScientific:2026ifv}. In either case, the common goal is to distinguish astrophysical signals from instrumental and environmental noise. The pipeline outputs consist of both significance statistics and (for the template-based searches) source-related quantities such as the masses of the merging compact objects. 

To quantify significance, the pipelines assign each candidate a \ac{FAR}: a right-sided p-value normalised by time, measuring how often noise alone would produce an equally significant event by chance.
Practically, this computation involves defining a test statistic $\rho$, empirically estimating the distribution of $\rho$ under the null hypothesis \citep[see][]{LIGOScientific:2026sit, LIGOScientific:2026ifv}, and then, for a candidate event with a measured test statistic of $\rho^*$, the \ac{FAR} is 
\begin{equation}
    \rm{FAR}(\rho^*) = \frac{\sum_{i=1}^n {\bf 1}(\rho_i \geq \rho^*)}{T_{\rm{obs}}}\,,
\end{equation}
where the numerator represents the number of background events with detection statistic $\rho_i$ exceeding $\rho^*$, and $T_{\text{obs}}$ is the total effective background observation time \citep{Capano:2016uif}. 

The background is typically estimated by time-shifting data streams between detectors, ensuring any surviving coincidences are of terrestrial origin \citep{LIGOScientific:2026ifv}. The exact form of the detection statistic varies between pipelines, with some using the matched-filter \ac{SNR} and others a more general ranking statistic \citep{LIGOScientific:2026ifv}.
Candidates with a \ac{FAR} below a given threshold, usually 1 per year, are classified as significant astrophysical signals \citep{LIGOScientific:2026ifv}.

Within the \ac{LVK} collaboration, there are multiple different search pipelines currently being routinely used. In this work we will consider four of them, due to data availability in the datasets we consider: \gstlal \citep{Messick:2016aqy, Sachdev:2019vvd, Tsukada:2023edh, Sakon:2022ibh, Joshi:2025nty,  Cannon:2020qnf, Hanna:2019ezx}, \mbta \citep{Allene:2025saz, Aubin:2020goo, Adams:2015ulm}, \pycbc \citep{Allen:2005fk, DalCanton:2014hxh, Usman:2015kfa, Nitz:2017svb, Allen:2004gu, Davies:2020tsx}, and \cwb \citep{Mishra:2024zzs, Klimenko:2005xv, Klimenko:2008fu, Klimenko:2015ypf, Klimenko:2004qh, Klimenko:2011hz}. The former three rely on matched filtering, while the latter performs an unmodelled, coherence-based search. Additional details are described in \citet{LIGOScientific:2026ifv}. 

The use of multiple pipelines introduces challenges when their assessments of a given candidate are inconsistent, for example, due to differences in methodologies and configuration choices or variations in sensitivity across the parameter space. It is therefore essential to combine the outputs of these pipelines into a single robust significance measure for each candidate.

In current \ac{LVK} analyses, this is typically achieved by comparing the \ac{FAR} across pipelines, and considering candidates as significant astrophysical signals if at least one pipeline reports a \ac{FAR} below 1 per year~\citep{GWTC5}.

Since the \ac{FAR} is a right-sided p-value normalised by time, combining \acp{FAR} from multiple pipelines is analogous to the classical statistical problem of combining p-values.
Therefore, possible alternative combination approaches include Fisher's method \citep{fisher2005ra} or the harmonic mean; however, the computed \acp{FAR} are not independent (since they analyse the same data), and no joint background exists across all pipelines to calibrate the harmonic mean.
\ac{GW} astronomers therefore adopt a simple, pragmatic approach and use the minimum \ac{FAR} across all pipelines.
This is easy to implement and conservative in that it makes the strongest claim supportable by any single pipeline, without inflating significance through combination.

This strategy is simple and computationally efficient. However, discrepancies between pipeline outputs can be informative, as they capture complementary aspects of candidate events and can be exploited to extract additional information. There are now several alternative approaches to combine pipelines, capturing correlations, see, for example \citet{Banagiri:2023ztt}, \citet{Tsukamoto:2025vuu}, and \citet{Ashton:2025jhn}. 

\begin{figure}[h!]
    \centering
    \includegraphics[width=0.95\textwidth]{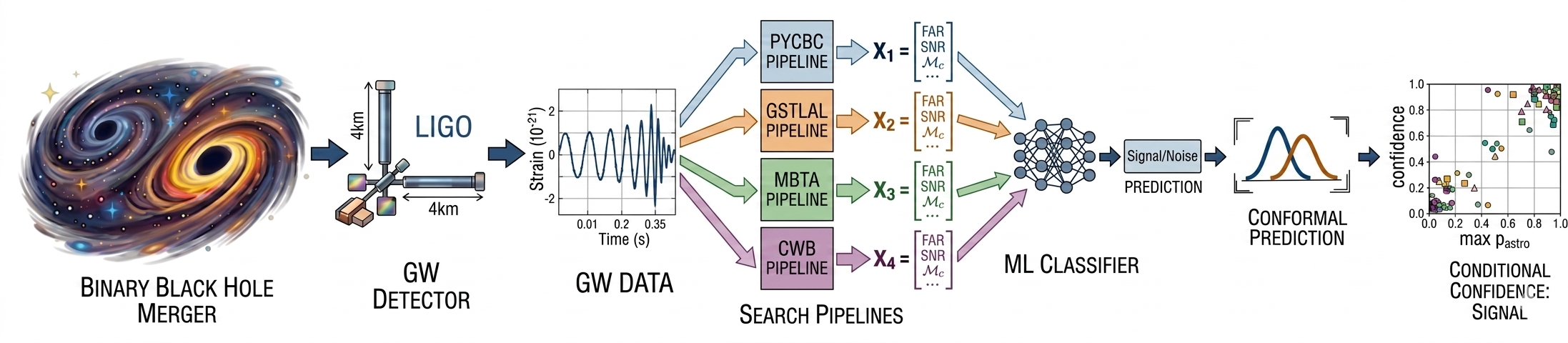}
    \caption{Illustration of the workflow from astrophysical event to our search pipeline combination framework. Two merging compact objects, such as black holes, generate gravitational waves, which are detected by the \acl{LVK} interferometers. The data is then analysed by several search pipelines, which output data products such as the \ac{FAR} and \ac{SNR}. This serves as input to our framework, applying an \ac{ML} classifier and \ac{CP} to combine pipeline outputs into a single conditional confidence value for each \ac{GW} candidate. The figure was created using Gemini 3.} 
    \label{fig:workflow}
\end{figure}

Of these, we focus on \citet{Ashton:2025jhn}, where we introduced a framework that uses a simple \ac{ML} model to learn correlations and combine outputs from multiple pipelines. Since \ac{ML} classifiers alone do not provide calibrated uncertainty estimates, we augment the framework with \ac{CP}, which provides statistically rigorous, distribution-free prediction sets with guaranteed coverage \citep{vovk2005algorithmic, angelopoulos2021gentle}. 
To provide astronomers with a single interpretable significance measure, we use the \textit{conditional confidence} defined in \citet{Ashton:2024wae}, an extension of the \ac{CP} confidence \citep{Shafer2007CP} that can always be efficiently computed in practice.
The conditional confidence quantifies how signal-like a candidate appears, and can be interpreted as the largest error rate for which the signal label remains admissible.
The resulting conditional confidence measures are essential for trustworthy astrophysical interpretation, as a lack of quantified uncertainties can lead to unreliable scientific conclusions.
\cref{fig:workflow} illustrates the workflow, from astrophysical event to the conditional confidence output by our pipeline combination framework. 

A central requirement for deploying such a method in practice is that it can be trained and calibrated on simulated mock data and then applied to real detector data, where the labels are unknown. 
However, existing mock data do not fully reproduce the complexity of real observations as they are constructed to stress-test the detection infrastructure rather than to reproduce accurate astrophysical distributions.
For example, detector noise characteristics, pipeline behaviour, and astrophysical signal properties all differ from reality. In particular, signal rates and source-class fractions are not representative of what has actually been observed, injected mass and spin distributions are deliberately broader and more generic than inferred astrophysical populations, and a minimum-SNR cutoff on injections excludes low-SNR events that would still be present in the true signal distribution even if not currently detectable~\citep{Chaudhary:2023vec, LLPIC:2026prep}. 
Such mismatches can lead to distribution shifts between calibration and testing datasets that invalidate the coverage guarantees of standard \ac{CP}.

We model this discrepancy between datasets as a distributional shift in the pipeline outputs. A full characterisation would require estimating how the feature distributions shift for each class separately, which is not tractable without labels from the test dataset. We therefore consider several approximations to the class-conditional ratio, ranging from the marginal feature ratio estimated from unlabelled test data to heuristic approaches based on pseudo-labelled test samples, as described in \cref{sec:method}. The marginal approximation is motivated by the observation that distributional shifts in pipeline outputs are expected to affect signal and noise events similarly, implying that the class-conditional and marginal ratios are approximately equal. 

To correct for this shift, we apply the weighted \ac{CP} method of \citet{tibshirani2019conformal}, which reweights calibration points by the estimated likelihood ratio of the test to calibration feature distributions, restoring the coverage guarantee of standard \ac{CP} without requiring retraining of the underlying classifier. 

We first induce controlled covariate shifts of varying intensity in our mock data to systematically investigate how distributional differences affect coverage and conditional confidence estimates, and how weighted \ac{CP} can mitigate their impact. We then apply the same framework to investigate the naturally occurring shift between two different mock datasets, using the controlled results as a reference to interpret and contextualise the observed behaviour in our astrophysical setting.

In this work, we apply the likelihood reweighting method of \citet{tibshirani2019conformal} to \ac{GW} search pipeline combination using Mondrian (label-conditional) binary classification. Applying weighted \ac{CP} to Mondrian classification is not novel in itself, e.g. \citet{laghuvarapu2026kmm} take a similar approach under covariate shift. However, we are not aware of any existing work explicitly addressing label-conditional covariate shift. Since exact correction is intractable without test labels, our methodological contribution is to introduce and empirically compare several practical approximations, ranging from marginal to pseudo-label-based approaches, as described in \cref{sec:method}. The primary novelty of this work, however, lies in the application to \ac{GW} astronomy, where we demonstrate that weighted \ac{CP} can correct for naturally occurring distribution shifts between datasets, with direct implications for the reliability of astrophysical conclusions.

\section{Method} \label{sec:method}
In this section, we describe the details of our \ac{GW} pipeline combination framework, following the same approach as in \citet{Ashton:2025jhn}. We take as input the open pipeline outputs produced by the \ac{LVK} collaboration, combine them using a \ac{ML} classifier, and subsequently apply \ac{CP} to quantify the uncertainty and produce well-calibrated conditional confidence scores for each candidate. 
We then describe the extension to weighted \ac{CP}~\citep{tibshirani2019conformal} to correct for distribution shift between the calibration and test distributions.

To train the classifier and calibrate \ac{CP}, we use labelled (signal or noise) mock datasets constructed from simulated signals injected into detector noise and processed by each of the search pipelines. 
In this work, we consider two such datasets: the pre-fourth-observing-run \ac{MDC}~\citep{Chaudhary:2023vec} and the \ac{LLPIC} dataset~\citep{LLPIC:2026prep}, produced during the fourth observing run. 
The \ac{LVK} detectors operate in coordinated observing runs, separated by commissioning periods during which both the detectors and analysis software (including the search pipelines) are upgraded, resulting in evolving data properties between runs.
The \ac{MDC} and \ac{LLPIC} datasets are based on replayed data from the third and fourth observing runs, respectively, and therefore reflect differences in detector configuration and pipeline implementations between runs. 
The \ac{MDC} dataset spans 40 days of data and contains 9948 candidates, of which 5910 are signals. 
In contrast, the \ac{LLPIC} dataset we use covers 7 days and contains 1204 candidates, including 534 signals. 
We split the \ac{MDC} data into training (80\%), calibration (10\%), and testing (10\%) subsets, while reserving the entire \ac{LLPIC} dataset for testing only. 

We restrict our inputs to the reported \ac{FAR} and \ac{SNR} from each of the four pipelines. We use the logarithm of the \aclu{IFAR} (\ac{IFAR}; \ac{IFAR}=1/\ac{FAR}) in place of the \ac{FAR}, as differences in statistical significance are meaningful on logarithmic scales. This choice also ensures that missing pipeline outputs, which we replace with zero, remain well defined.

In this work, we use a \ac{LR} classifier due to its simplicity and interpretability, see \citet{Malz:2026nmi} for discussion on different classifiers.
The \ac{ML} model is trained on sample–label pairs, $(\vec{X}, y)$, where $\vec{X}$ denotes the feature vector composed of the output from the pipelines and $y \in \{0,1\}$ the corresponding label (0 for noise, 1 for signal). \ac{LR} models the probability of a candidate being a signal as a sigmoid function of a linear combination of the input features, optimised via the binary cross-entropy loss. We then use the output probability of the trained classifier to define the nonconformity measure for \ac{CP}.

\subsection{Conformal prediction}
We employ Mondrian (label-conditional) \ac{CP}~\citep{vovk2013conditional, ding2023CP} to guarantee conditional coverage for each label, which is essential for our \ac{GW} analysis to ensure reliable uncertainty quantification for both signal and noise events. Applying \ac{CP} for binary classification, we use the classifier output probability $p_y(\vec{X})$ for each label $y$ and define the nonconformity score 
\begin{equation}
    s^y(\vec{X}) = 1- p_y(\vec{X})\,.
    \label{eq:s}
\end{equation}
We compute the calibration scores, $s^y(\vec{X}_i)$, $y=0, 1$, for each calibration sample $\vec{X}_i$, group them by class label $y$, and within each class, sort them in increasing order. 
The corrected $(1-\alpha)$-th empirical quantile is then estimated as
\begin{equation}
    \hat{q}_\alpha^y = s^y_{\lceil (N_y+1)(1-\alpha) \rceil}\,,\quad N_y = \sum_{i=1}^N \mathbf{1} (Y_i=y)\,,
    \label{eq:qhat}
\end{equation}
where $\alpha \in [0,1]$ is the user-chosen error rate, and $N_y$ is the number of calibration samples with label $y$, which, in general, may be different from $\frac N2$.
For a new test sample $\vec{X}'$, we compute the nonconformity scores $s^y(\vec{X}')$ for all $y =0, 1$, and let the prediction set be
\begin{equation}
    \Gamma^\alpha(\vec{X}') = \{ y \in \{0, 1 \} : s^y(\vec{X}')\le \hat{q}_\alpha^y \}\,.
\end{equation}
\ac{CP} guarantees finite-sample validity, ensuring that the true test label ${Y'}$ satisfies
\begin{equation}
    1-\alpha\leq\textrm{Pr}({Y'}\in\Gamma^\alpha(\vec{X'})|Y'=y)\leq 1-\alpha+\frac{1}{N_y+1}\,.
    \label{eq:cp_validity}
\end{equation}
For Mondrian \ac{CP}, these guarantees hold conditionally within each class, in addition to marginal validity over the full dataset.

To determine the significance of each \ac{GW} candidate, we use the conditional confidence of the signal class.
Following \citet{Ashton:2024wae}, this quantity is defined as the value of $\alpha$ where the signal label becomes included in the prediction set $\Gamma^\alpha$. Formally,
\begin{equation}
    \texttt{conf}_{Y=1}(\vec{X}) = \max \left\{ \alpha \in [0,1] : 1 \in \Gamma^{\alpha}(\vec{X}) \right\} = \max \{\alpha \in [0, 1]: 1-p_1(\vec{X}) \leq q^1_\alpha \}\,.
\end{equation}
Defining a threshold on $\texttt{conf}_{Y=1}$ (e.g. at $0.5$) yields a catalogue with controlled purity.

\subsection{Conformal prediction under distribution shift}
Standard \ac{CP} relies on the assumption of exchangeability between calibration and test data, which is violated when they come from different distributions, $P^{\textrm{cal}}_{XY} \neq P^{\textrm{test}}_{XY}$. 
A special case of distribution shift is covariate shift, where the marginal distribution of features differs between calibration and test data but the conditional distribution of the labels given the features remains unchanged, that is, $P^{\textrm{cal}}_{XY} = P^{\textrm{cal}}_{X} P^{\textrm{cal}}_{Y|X} \neq P^{\textrm{test}}_{X} P^{\textrm{cal}}_{Y|X} = P^{\textrm{test}}_{XY} $.
\citet{tibshirani2019conformal} proposed to correct such a covariate shift by reweighting each calibration point with the likelihood ratio between test and calibration distributions.

The scheme does not apply straightforwardly to our setup, which requires conditioning on the labels.
The distribution shift we need to address is
\begin{equation}
    P^{\textrm{test}}_{X|Y=y} \neq P^{\textrm{cal}}_{X|Y=y}, \quad y=0, 1 \,.
\end{equation}
As $P_{X|Y}=P_{Y|X} \frac{P_X}{P_Y}$, such a \emph{conditional covariate shift} is implied by any non-trivial covariate shift ($P_{Y|X}$ invariant, $P_X$ shifting).

Let the conditional calibration and test distributions be $P_{X|Y}$ and $\tilde{P}_{X|Y}$.
For each $y=0, 1$, the \emph{conditional likelihood ratio} is then
\begin{equation}
    w_y(\vec{X}) = \frac{\mathrm{d} \tilde{P}_{X|Y=y}(\vec{X})}{\mathrm{d} P_{X|Y=y}(\vec{X})}\,.
\end{equation}
The idea is to weight up the calibration points whose covariates resemble the test points and down-weight those that do not. 
Similar to \citet{tibshirani2019conformal}, reweighting the calibration samples fully restores validity when the likelihood ratios are known.
On the one hand, this is remarkable: general distribution shifts could be resolved by combining sample reweighting and Mondrian (label-conditional) CP.
As expected, however, this is impossible in any real-world scenarios where conditional likelihood ratios $w_y(\vec{X})$ are unknown and must be estimated (often by training a classifier that discriminates between calibration and test samples).
Unlike the marginal likelihood ratios of \citet{tibshirani2019conformal}, which can be learned consistently from unlabelled covariate data, defining an unbiased estimator of the conditional likelihood ratios is more challenging, because it would require knowing the test labels.

In non-conditional likelihood ratio estimation, we assign calibration and test points a set-class label, $C=0, 1$, and train a probabilistic classifier, $\hat{p}_C(\vec{X})$, (we use a simple random forest as in \citet{tibshirani2019conformal}) on the combined set. 

For conditional ratios, the procedure should be performed separately on the calibration and test samples with labels $y=0$ and $y=1$.
Let $\hat{p}_{y, C}(\vec{X})$, $y=0, 1$, be the binary classifiers trained for $y=0$ and $y=1$.
The corresponding estimated likelihood ratio is
\begin{equation}
    \hat{w}_y(\vec{X}) = \frac{\hat{p}_{y, C=1}(\vec{X})}{1 - \hat{p}_{y, C=1}(\vec{X})}\,,
    \label{eq:weight_estimation}
\end{equation}
where $\hat{p}_{y,C=1}(\vec{X}) = \hat{p}(C=1 \mid \vec{X}, Y=y)$ is the estimated probability of a point belonging to the distribution of the test points with label $y$.
Without knowing the test labels, selecting the test data needed to train the label-conditional weight estimator is not possible. We therefore approximate the conditional likelihood ratio $w_y(\vec{X}) \approx \tilde{w}_y(\vec{X})$ using one of the following strategies:
\begin{itemize}
    \item \textbf{Marginal}: estimate $\tilde{w}_y(\vec{X}) = \frac{d\tilde{P}_{X}(\vec{X})}{dP_{X}(\vec{X})}$ using \emph{all} test samples and \emph{all} calibration samples.
    \item \textbf{Semi-conditional}: estimate $\tilde{w}_y(\vec{X}) = \frac{d\tilde{P}_{X}(\vec{X})}{dP_{X|Y=y}(\vec{X})}$ using \emph{all} test samples and only the calibration samples \emph{with label $y$}.
\end{itemize}
We also consider two heuristic approaches based on pseudo-labelling the test 
samples:
\begin{itemize}
    \item \textbf{Classifier conditional}: estimate $\tilde{w}_y(\vec{X}) = \frac{d\tilde{P}_{X|\hat{Y}=y}(\vec{X})}{dP_{X|\hat{Y}=y}(\vec{X})}$, where $\hat{Y} = \mathbf{1}(p_1(\vec{X}) \geq \tfrac{1}{2})$ is the label predicted by the signal classifier.
    \item \textbf{\maxifar conditional}: estimate $\tilde{w}_y(\vec{X}) = \frac{d\tilde{P}_{X|\tilde{Y}=y}(\vec{X})}{dP_{X|\tilde{Y}=y}(\vec{X})}$,  where $\tilde{Y} = \mathbf{1}(\max\log_{10}\rm{IFAR} \gtrsim 7.5)$ is a proxy label based on the conventional \ac{FAR} threshold of 1 per year.
\end{itemize}

Given one of these approximations, we denote the resulting estimated likelihood ratio as $\hat{w}(\vec{X})$ and proceed as in \citet{tibshirani2019conformal}, applying $\hat{w}$ to both calibration and test data to obtain weights $w_{\textrm{cal}}(\vec{X}_i) = \hat{w}(\vec{X}_i)$ and $w_{\textrm{test}}(\vec{X}') = \hat{w}(\vec{X}')$, respectively.

The standard empirical quantile in \cref{eq:qhat} is replaced by a weighted quantile. For a test point $\vec{X}'$, the normalised weights of the calibration points with label $y$, are
\begin{equation}
    p_i^{w, y}(\vec{X}') = \frac{w_{\textrm{cal}}(\vec{X}_i)}{\sum_{j=1}^{N_y} w_{\textrm{cal}}(\vec{X}_j) + w_{\textrm{test}}(\vec{X}')},
    \quad i = 1, \ldots, N_y,
\end{equation}
and the weight for the test point is
\begin{equation}
    p_{\textrm{test}}^{w,y}(\vec{X}') = \frac{w_{\textrm{test}}(\vec{X}')}{\sum_{j=1}^{N_{y}} w_{\textrm{cal}}(\vec{X}_j) + w_{\textrm{test}}(\vec{X}')}\,,
\end{equation}
where $i = 1,\ldots,N_{y}$ indexes calibration points with label $y$. 
The per-label weighted quantile threshold is then
\begin{equation}
    \hat{q}_\alpha^{w, y}(\vec{X}') = \mathrm{Quantile}\!\left(1-\alpha;\; \sum_{i=1}^{N_y} p_i^{w,y}(\vec{X}')\, \delta_{s_i^y} + p_{\textrm{test}}^{w,y}(\vec{X}')\, \delta_\infty \right)\,,
    \label{eq:qhat_weighted}
\end{equation}
where $\delta_a$ denotes a point mass at $a$ (the distribution assigning all of its probability to the value $a$), following \citet{tibshirani2019conformal}, and the argument is a weighted discrete probability distribution over the nonconformity scores, where each score $s_i^y$ gets weight $p_i^{w,y}$ and $\infty$ gets weight $p_{\textrm{test}}^{w,y}$ (the $\infty$ atom is a standard conformal trick to account for the test point having an unknown score). 
The weighted prediction set is then defined as
\begin{equation}
    \Gamma^\alpha_w(\vec{X}') = \left\{ y \in \{0, 1\} : s^y(\vec{X}') \le \hat{q}_\alpha^{w, y}(\vec{X}') \right\}\,.
\end{equation}
Thus, the weighted quantile, $\hat{q}_\alpha^{w, y}(\vec{X}')$, depends on the test point, $\vec{X}'$, through the sample weights, and must be recomputed for each new test sample. 

Similar to \citet{tibshirani2019conformal}, if oracle conditional ratios, $\bar w_{y}(\vec{X})$, $y=0, 1$, were known (or consistently estimated), nominal coverage guarantees would hold as in \cref{eq:cp_validity}. That is, 
\begin{equation}
    \mathrm{Pr}\!\left({Y'} \in \Gamma^\alpha_{\bar w_y} (\vec{X}')  \mid {Y}'=y\right) \geq 1 - \alpha\,.
    \label{eq:cp_validity_weighted}
\end{equation}

Even in this ideal case, the finite-sample bound is less tight than in  Eq.~\eqref{eq:cp_validity} because the effective number of calibration samples contributing to the quantile estimate is reduced. Following \citet{tibshirani2019conformal}, we compute the effective sample size \citep{kish1965survey} for weighted \ac{CP} as
\begin{equation}
    \hat{n}_y = \frac{\left(\sum_{i=1}^{N_y} w_{\textrm{cal}}(\vec{X}_i)\right)^2}{\sum_{i=1}^{N_y} w_{\textrm{cal}}(\vec{X}_i)^2}\,.
\end{equation}
Intuitively, $\hat{n}_y$ measures how many equally-weighted calibration points would yield equivalent precision in the quantile estimate. A highly concentrated test distribution (relative to the calibration distribution) will therefore result in a small $\hat{n}_y$, increasing the variance of the coverage.

Unlike the marginal covariate shift setting of \citet{tibshirani2019conformal}, however, the conditional ratios $w_y(\vec{X}) = d\tilde{P}_{X \mid Y=y}/dP_{X \mid Y=y}$ cannot be consistently estimated without knowing the test labels, which would make the problem trivial. The approximations introduced above inevitably induce a bias in the weight estimates that cannot be bounded without further assumptions, and whose impact on coverage depends on the data. Our contribution is therefore to reduce the general problem of \ac{CP} under unconstrained distribution shift in classification to a practical one: training a weight estimator classifier under approximate or partly unknown labels, and investigating empirically how well coverage is preserved. Providing theoretical guarantees under reasonable assumptions on the data distribution, for example, by bounding the divergence between $P_{X \mid Y=0}$ and $P_{X \mid Y=1}$, remains an important direction which we leave for future work.

\section{Experiments} \label{sec:results}
To investigate the impact of covariate shift and the effectiveness of weighted \ac{CP} on our \ac{GW} data, we run two experiments: first, we construct controlled distributional shifts in the \ac{MDC} test data and, second, we explore the existing shift between the \ac{MDC} and \ac{LLPIC} datasets. 

\subsection{Controlled covariate shift}
We split the \ac{MDC} dataset into training, calibration, and test sets as described in \cref{sec:method}, and induce covariate shift by resampling the test set according to feature-dependent weights, preserving $P_{Y \mid X}$ while altering the marginal distribution $P_X$. 

Simulating a scenario in which one of the pipelines exhibits distributional changes between mock and real data, 
we apply reweighting to the \ac{IFAR} and \ac{SNR} features of the selected pipeline, while leaving the features from other pipelines unchanged. 

Since the \ac{IFAR} and \ac{SNR} are strongly correlated, we only consider the case where both features have the same weighting. Covariate shift is introduced by upweighting samples whose feature values lie within a specified percentile range, thereby biasing the resampled test distribution toward higher or lower regions of the feature space. The magnitude of this perturbation is governed by a multiplicative scaling parameter on the weights, which we refer to as the shift intensity. 

We further restrict this setup by considering two types of shifts: one targeting lower feature values (resampling within the 10th–30th percentiles) and one targeting higher feature values (70th–90th percentiles). For each case, we vary the shift intensity over the values 2, 5, and 10, where a value of 1 corresponds to no shift, as it leaves the sampling distribution unchanged.

\begin{figure}[h!]
    \centering
    \includegraphics[width=0.75\textwidth]{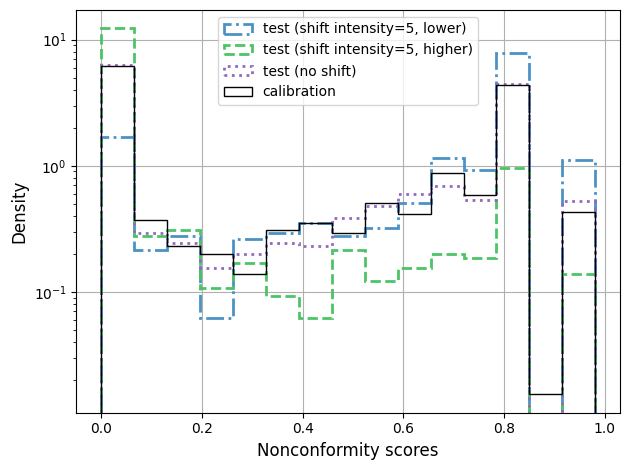}
    \caption{Histogram showing the values of the nonconformity scores of the signal class for the calibration data, unshifted test data and two sets of shifted test data with shift intensity 5, resampled to higher and lower feature values, respectively. } 
    \label{fig:nonconf_hist}
\end{figure}

To illustrate the distribution shift, we plot the histogram of the nonconformity scores in \cref{fig:nonconf_hist}, for two of the defined test cases together with those for the calibration data and the unshifted test data. As expected, the nonconformity scores of the calibration and unshifted test data agree well. The test set shifted to lower feature values results in higher nonconformity scores compared to the calibration and unshifted test, while the opposite holds for the test data shifted to higher feature values. 

We quantify the induced covariate distribution shifts using the \ac{EMD}, which measures the distance between two distributions as the minimal cost of transporting probability mass from one to the other \citep{rubner1998metric}. 
Formally, the \ac{EMD} between the calibration distribution $P_X$ and the test distribution $\tilde{P}_X$ is defined as the first Wasserstein distance,
\begin{equation}
    \text{EMD}(P_X, \tilde{P}_X) = \inf_{\pi \in \Gamma(P_X, \tilde{P}_X)} 
    \int_{\mathbb{R} \times \mathbb{R}} |z - z'| \, \mathrm{d}\pi(z, z')\,,
\end{equation}
where $\Gamma(P_X, \tilde{P}_X)$ is the set of joint distributions on $\mathbb{R} \times \mathbb{R}$ with marginals $P_X$ and $\tilde{P}_X$, and where $z$ and $z'$ are scalar feature values drawn from $P_X$ and $\tilde{P}_X$ respectively~\citep{ramdas2017wasserstein}. In practice, we compute this using the \texttt{scipy.stats.wasserstein\_distance}~\citep{virtanen2020scipy} implementation, applied feature-wise and averaged across all input features.

Larger \ac{EMD} values thus correspond to stronger shifts. 
The resulting average \ac{EMD} values for each shift setting are reported in \cref{tab:emd}. 
The values demonstrate how a stronger shift intensity results in a higher \ac{EMD}, and how resampling the test data in the region of higher feature values gives a slightly greater shift compared to resampling in lower feature values with the same shift intensity. The asymmetry between the lower and higher shift cases at the same intensity reflects the shape of the underlying feature distribution: shifting toward higher feature values moves the test distribution further into the tail, where the calibration density is low, resulting in a greater distributional distance. 

\begin{table}
\floatconts
  {tab:emd}
  {\caption{\ac{EMD} values, averaged over all features, quantifying the strength of the distribution shifts between the shifted test datasets compared to the calibration dataset. For each shift intensity, we resample in lower and higher regions of feature space, respectively, obtaining six shifted test sets. For comparison, the \ac{EMD} value between the unshifted test and the calibration distributions is $0.040$.}\vspace{-1.5em}}
  {\begin{tabular}{ccc}
  \toprule
  \bfseries Shift intensity & \bfseries lower & \bfseries higher \\
  \midrule
    2 & 0.22 & 0.17 \\
    5 & 0.36 & 0.43 \\
    10 & 0.40 & 0.55 \\
  \bottomrule
  \end{tabular}}
\end{table}

\begin{figure}[h]
    \centering
    \subfigure[Lower]{\includegraphics[width=0.45\textwidth]{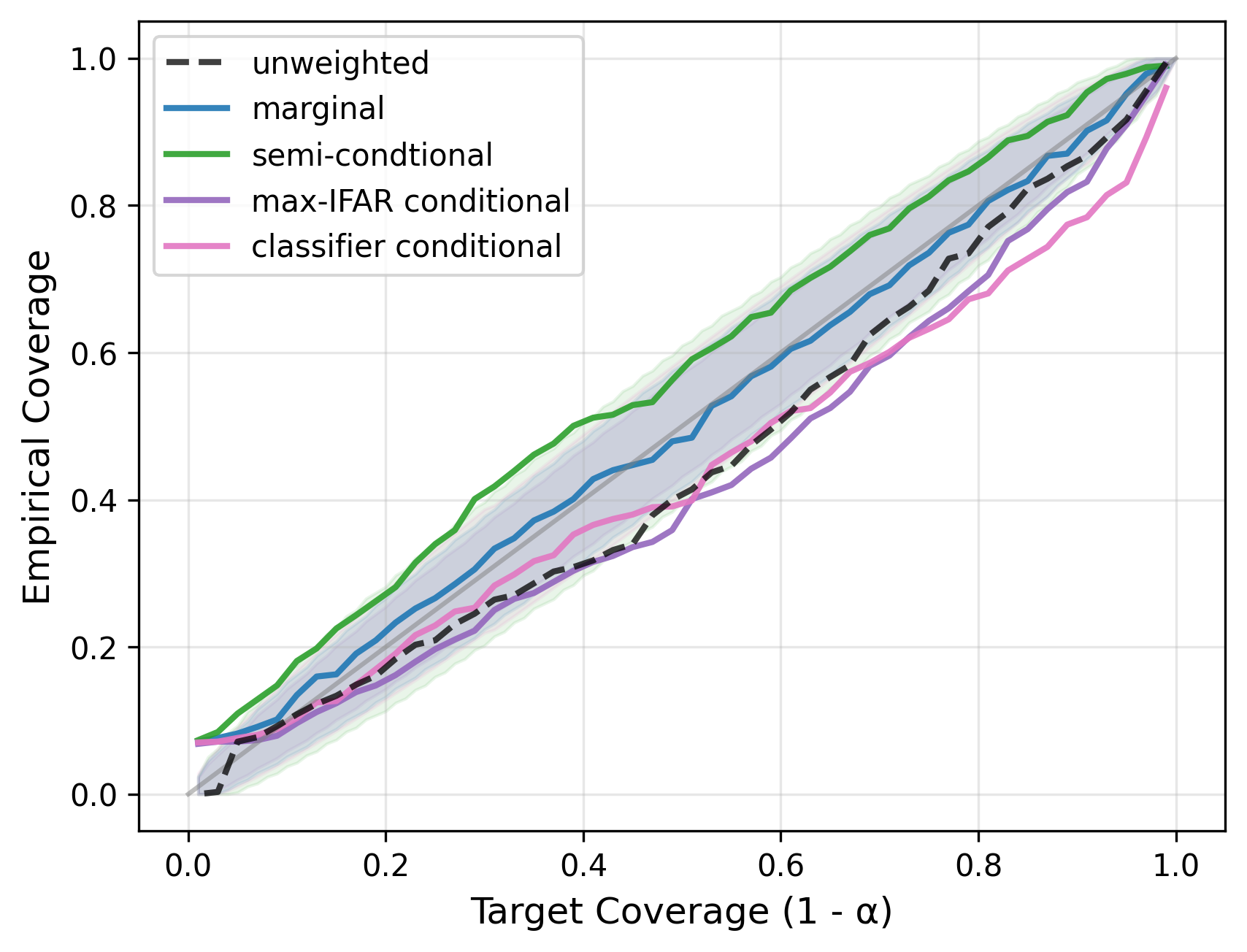} \label{fig:coverage_shifted_lower}}
    \subfigure[Higher]{\includegraphics[width=0.45\textwidth]{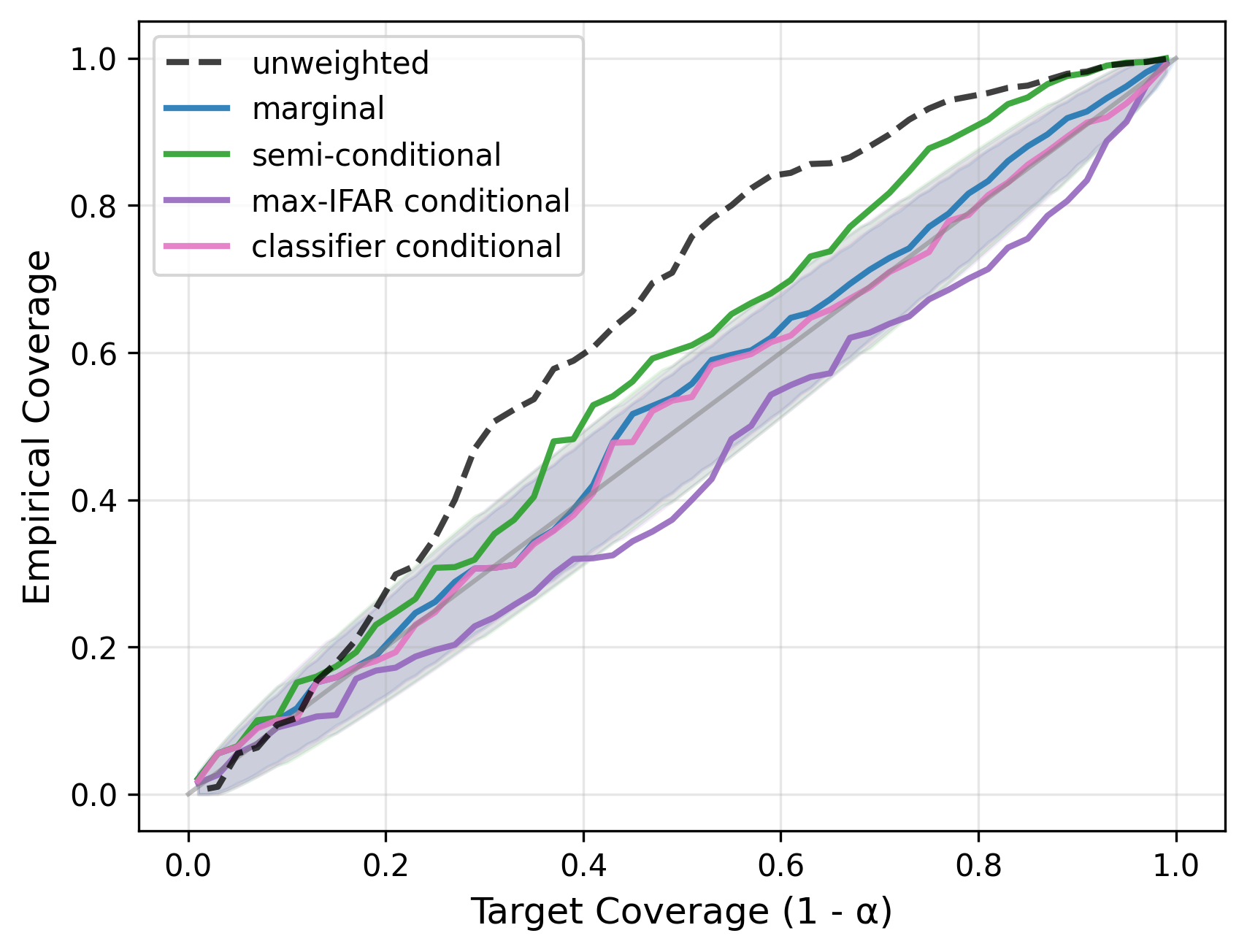} \label{fig:coverage_shifted_higher}}
    \caption{Empirical coverage comparing different \ac{CP} weighting methods, as obtained on the shifted test data for shift intensity 5, resampling features to lower (left) and higher (right) values, respectively. 
    The shaded areas represent the 99\% binomial confidence bands corresponding to the minimum per-label effective calibration size $\min_y \hat{n}_y$ for each weighting method, with wider bands corresponding to smaller effective sample sizes.} 
    \label{fig:coverage_method_comparison}
\end{figure}

Next, we compare the likelihood ratio approximations introduced in \cref{sec:method} empirically, applying each method to two of the defined test shifts and presenting the resulting coverage in \cref{fig:coverage_method_comparison}.
We observe that only the marginal likelihood ratio, where the weight estimator is trained on all test and all calibration samples, successfully recovers well-calibrated coverage in both test cases. 
Interestingly, when we induce overcoverage by shifting the features to higher values in \cref{fig:coverage_shifted_higher}, the approximate conditional likelihood ratio using the classifier prediction as the label of test data, also obtains valid coverage. 
However, methods that rely on a proxy for the true label, whether the classifier prediction or an alternative such as the \maxifar, introduce an additional source of error: any misclassification or inaccuracy in the proxy propagates into the weight estimates, and coverage guarantees are only recovered to the extent that the proxy is reliable. Since proxy quality cannot be guaranteed in general (as illustrated by \cref{fig:coverage_method_comparison}), and may degrade precisely in the regime where the distribution shift is most severe, and correction is most needed, we adopt the marginal likelihood ratio as the most robust and assumption-free choice. 
In the remainder of this work, we will apply the marginal weights only and refer to this as the \emph{weighted \ac{CP}} method. 

\begin{figure}[h]
    \centering
    \subfigure[Standard \ac{CP}]{\includegraphics[width=0.45\textwidth]{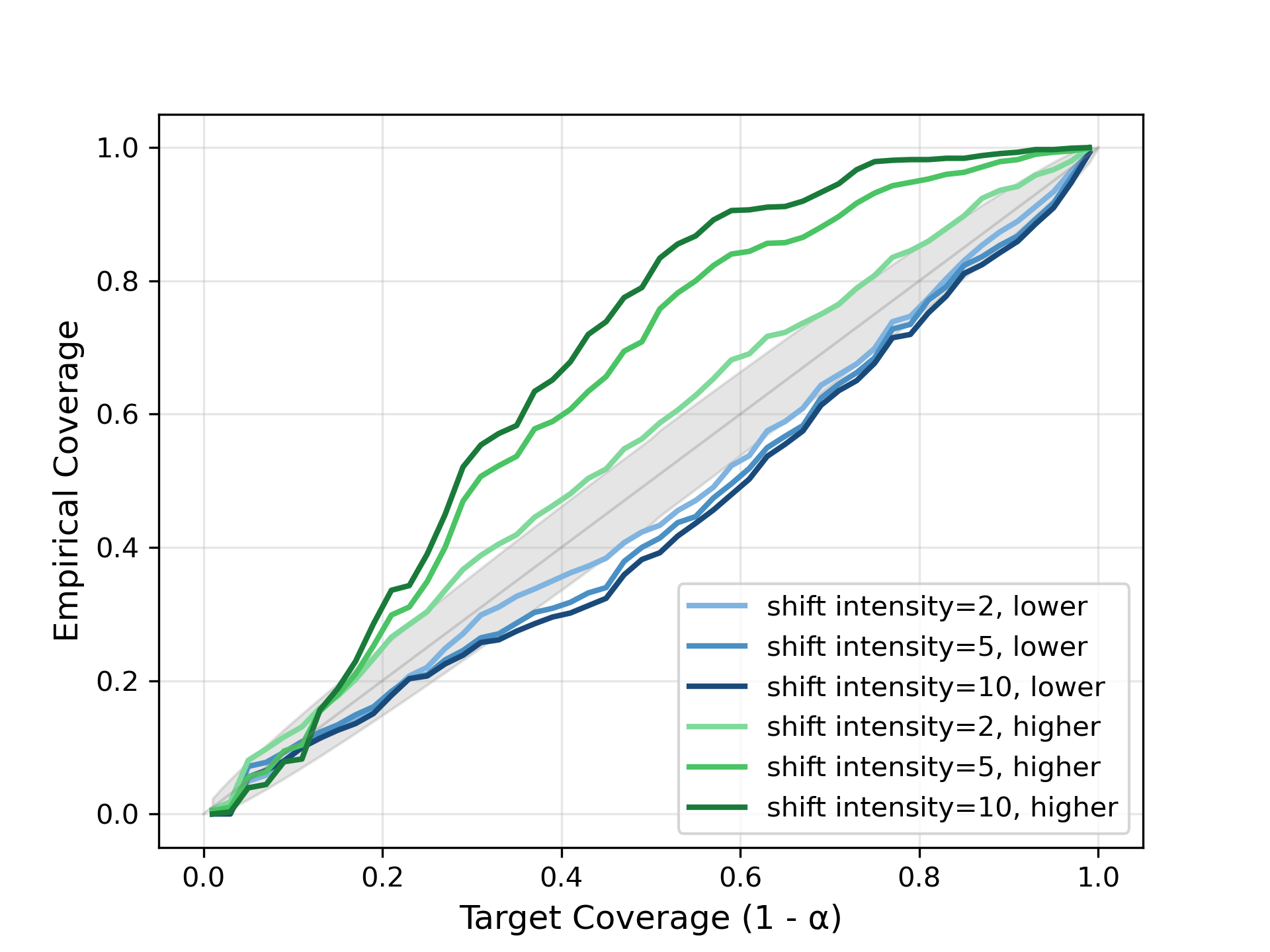} \label{fig:coverage_shifted}}
    \subfigure[Marginal weighted \ac{CP}]{\includegraphics[width=0.45\textwidth]{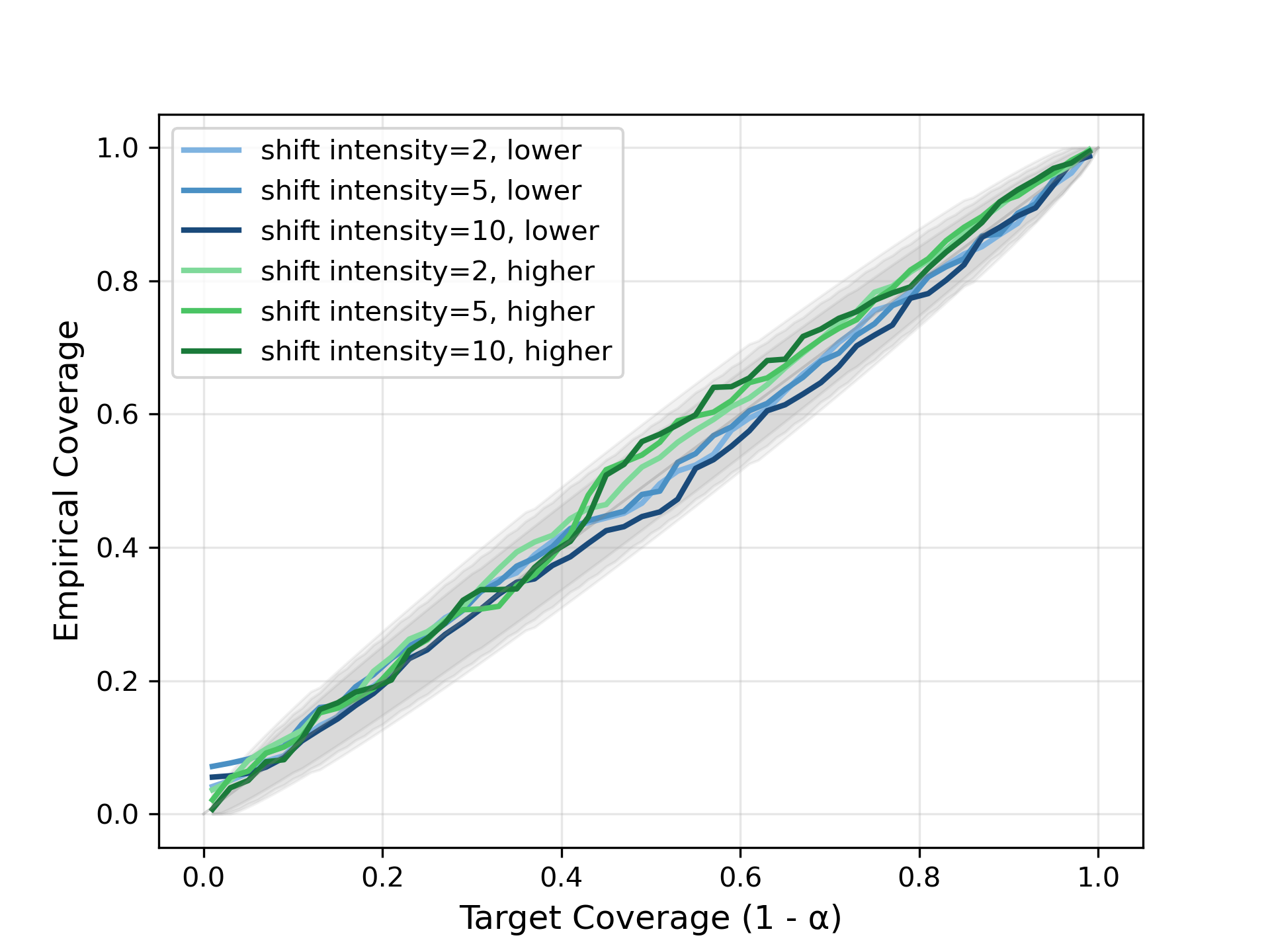} \label{fig:coverage_shifted_weighted}}
    \caption{Empirical coverage comparing standard Mondrian \ac{CP} (left) and marginal weighted Mondrian \ac{CP} (right), as obtained on the shifted test data for varying shift intensity.
    The shaded grey areas represent the 99\% binomial confidence bands corresponding to the minimum per-label effective calibration size $\min_y \hat{n}_y$ for each shift intensity, with wider bands corresponding to larger shift intensities. For standard \ac{CP} $\hat{n}_y$ reduces to $N_y$.} 
    \label{fig:coverage}
\end{figure}

Having defined and validated the experimental setup, we apply our framework to each of the shifted datasets, using both standard, unweighted, \ac{CP} and weighted \ac{CP}. For each method, we compute the empirical coverage, and report the results in \cref{fig:coverage}.

\cref{fig:coverage_shifted} shows the effect of the covariate shift on the coverage when using standard \ac{CP}. As expected, increasing the shift intensity moves the coverage curve further away from the diagonal. Shifting the test distribution toward lower feature values results in undercoverage, while shifting toward higher values leads to overcoverage. 
In our data, higher \ac{IFAR} and \ac{SNR} values indicate that an event is more likely to be a signal. 
Resampling the test distribution in the region of higher feature values increases the proportion of signal-like events, shifting the empirical distribution of test nonconformity scores downward, relative to the calibration set. Since the calibration quantile was computed from a different region of feature space, it becomes too permissive for the test set, resulting in overall larger average prediction sets and systematic overcoverage. 
Conversely, resampling toward lower feature values enriches the test set with noise-like events whose nonconformity scores are systematically higher, yielding overall smaller prediction sets and undercoverage.

Mondrian \ac{CP} enforces coverage validity separately within each class under the assumption of exchangeability. However, this guarantee does not extend to covariate shift that alters the marginal feature distribution $P_X$, and consequently $P_{X \mid Y}$, within each class, since the per-class calibration quantiles are still computed under the assumption that calibration and test features are exchangeable within each class. We note that when using non-Mondrian \ac{CP}, the under- and over-coverage obtained is more severe. 

\cref{fig:coverage_shifted_weighted} shows the empirical coverage obtained using weighted \ac{CP}. For all shift intensities, the coverage is approximately diagonal within the expected binomial noise, indicating that the weighted \ac{CP} returns well-calibrated prediction sets. 

To quantify the expected deviation from the diagonal in \cref{fig:coverage} due to finite sample size, we plot 99\% binomial confidence bands corresponding to the minimum per-label calibration size $\min_y N_y$ in \cref{fig:coverage_shifted}.
For the weighted \ac{CP} in \cref{fig:coverage_shifted_weighted}, the limiting uncertainty is the per-label effective calibration size $\hat{n}_y$, which decreases as the shift intensity increases and fewer calibration points contribute meaningfully to the quantile estimate. Thus, the expected deviation from the diagonal increases with shift intensity. Due to the symmetric nature of our set-up, $\hat{n}_y$ varies little between the lower and higher shift cases at the same shift intensity, and thus we only show one band per intensity level. 

We observe that while all shifted coverage curves are outside the shaded band in \cref{fig:coverage_shifted}, the curves are well within the expected deviation from the diagonal when using the weighted \ac{CP} in \cref{fig:coverage_shifted_weighted}, confirming that likelihood reweighting can successfully mitigate the covariate shift.

Having shown that the weighted \ac{CP} can successfully mitigate our covariate shift examples, we now investigate how covariate shift and subsequent weighted \ac{CP} affect the conditional confidence, as this is the output from our \ac{GW} search pipeline combination analysis we are most interested in. 

\begin{figure}[h!]
    \centering
    \includegraphics[width=0.75\textwidth]{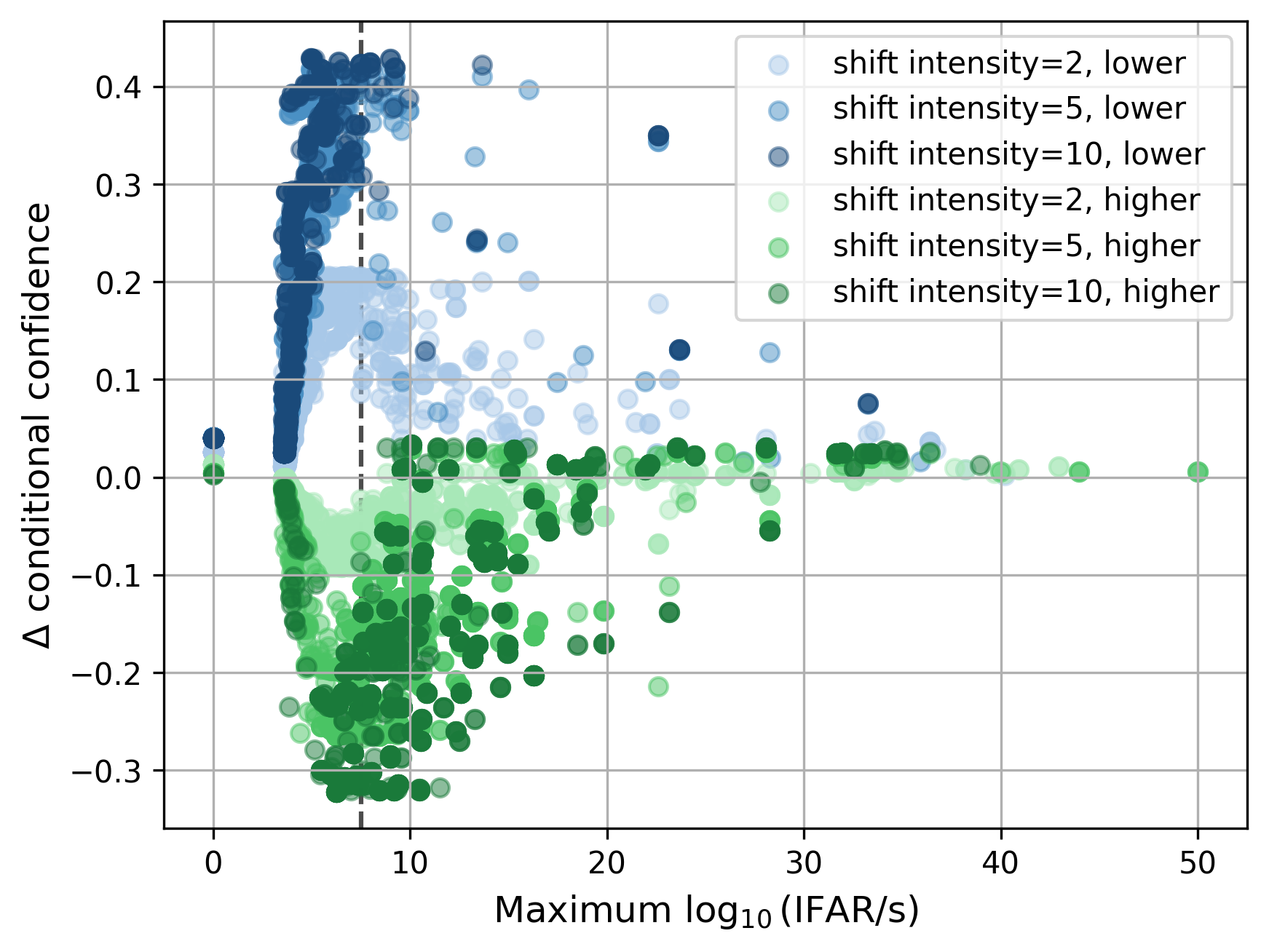}
    \caption{Difference between the conditional confidence as obtained using weighted \ac{CP} versus standard \ac{CP}. A positive difference implies that the conditional confidence increases when applying the weighted \ac{CP}. The dashed black vertical line indicates the \ac{FAR} threshold of 1 per year.} 
    \label{fig:confidence_residual}
\end{figure}

To investigate how the conditional confidence changes before and after applying the weighted \ac{CP}, we plot the conditional confidence difference between the two, defined as $\Delta \, \texttt{conf}_{Y=1} = \texttt{conf}_{Y=1}^{\hat{q}_\alpha^{w, y}(\vec{X}')} - \texttt{conf}_{Y=1}^{\hat{q}_\alpha^y}$, in \cref{fig:confidence_residual}. 
In the scenarios where features are shifted to lower values, we observe that the conditional confidence increases when using weighted \ac{CP} versus the standard \ac{CP}. The difference is also larger for a greater shift intensity. 
Conversely, when features are shifted to higher values, we observe that the conditional confidence decreases when using weighted \ac{CP} versus the standard \ac{CP}. In this scenario, there is also minor fluctuation around zero with some positive differences.

We plot the conditional confidence difference against the \maxifar, which represents a measure of candidate significance, to demonstrate the dependence on the underlying features. 
The conditional confidence varies the most for events with a \maxifar value close to the \ac{FAR} threshold, the conventional but arbitrary boundary between signal and noise, where classification is most uncertain. 
Meanwhile, for events with high \maxifar the conditional confidence barely changes, since these are clear signals with conditional confidence values around one. Similarly, at very low \maxifar values we find noise events with consistently small conditional confidence.

\subsection{Unknown shift between two datasets}
We now evaluate the performance of the \ac{MDC}-trained-and-calibrated framework on the \ac{LLPIC} dataset, simulating a situation where our pre-trained framework is later applied to real data. A distribution shift between the two mock datasets is expected, as they originate from different observing runs with associated upgrades to detectors and analysis software (changes which are also expected to contribute to the shift between mock and real data), although the nature of the shift is not known. 
Motivated by the expectation that distributional shifts in pipeline outputs affect signal and noise events similarly, and by our empirical findings in the previous section, we again adopt the marginally weighted \ac{CP} approach.

First, we quantify the shift using \ac{EMD}, obtaining a value of $0.22$ between the \ac{MDC} calibration data and the \ac{LLPIC} test set. This indicates a moderate shift, comparable to the induced covariate shifts with shift intensity 2 from \cref{tab:emd}. 

\begin{figure}[h!]
    \centering
    \subfigure[Coverage]{\includegraphics[width=0.45\textwidth]{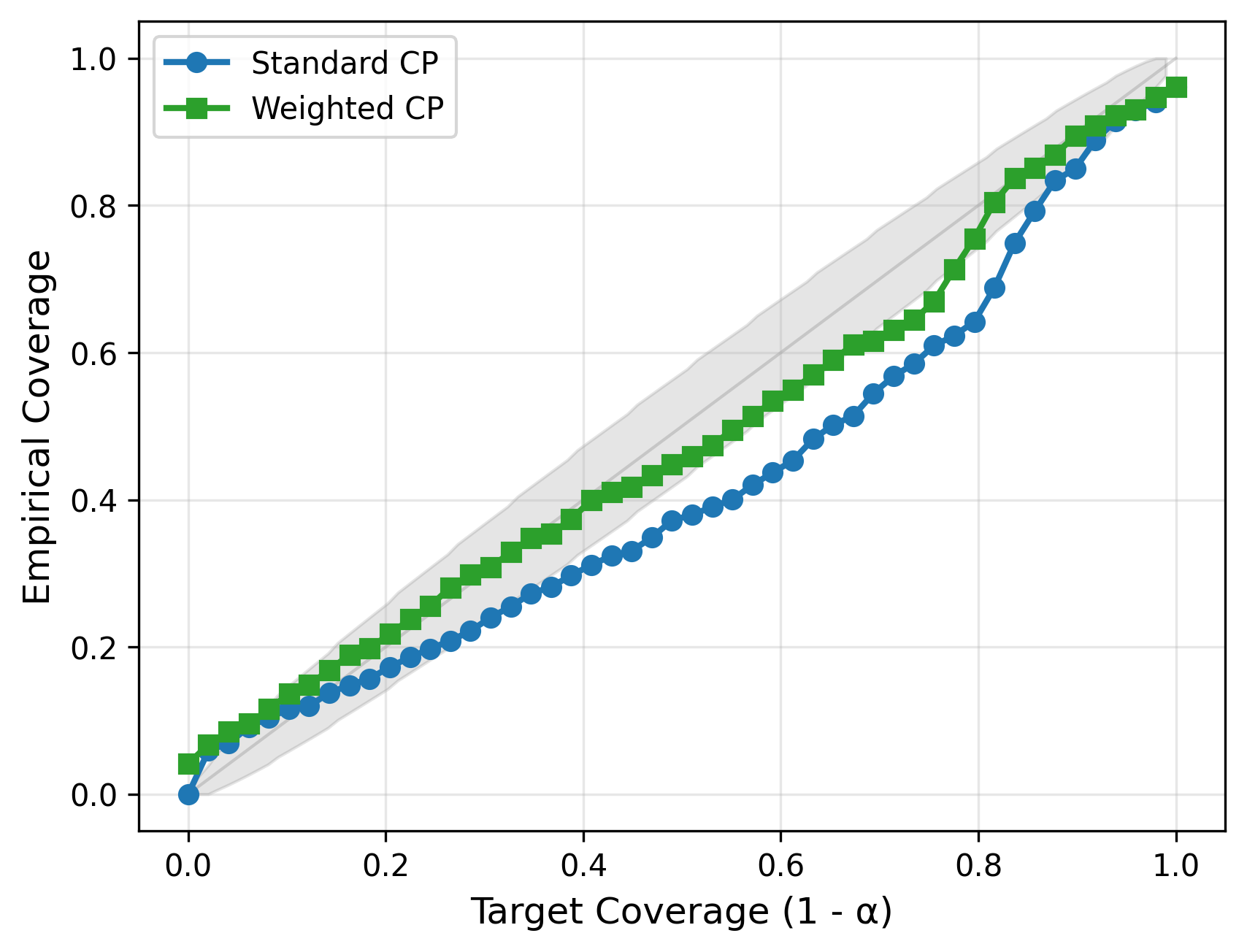} \label{fig:coverage_llpic}}
    \subfigure[Conditional confidence difference]{\includegraphics[width=0.45\textwidth]{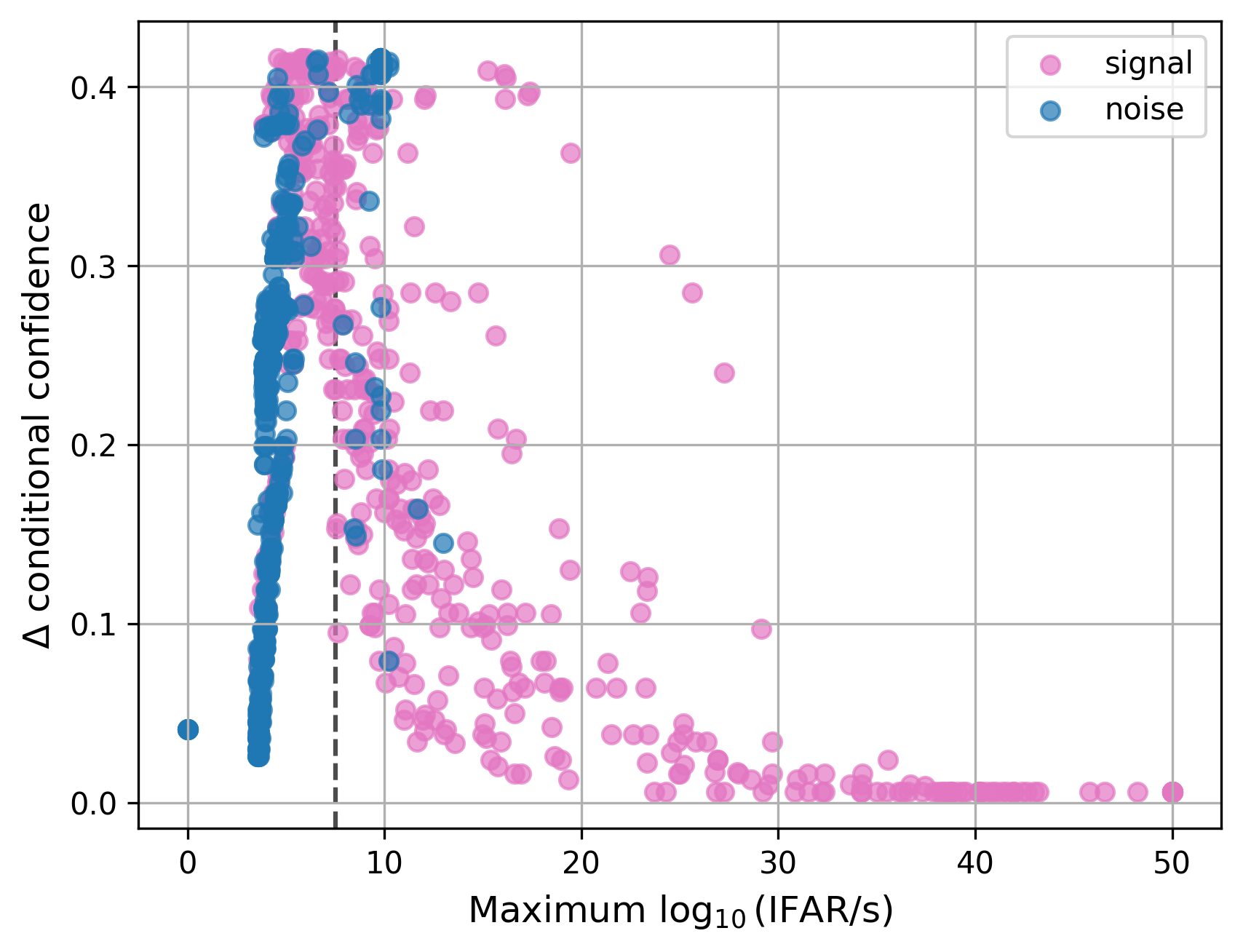} \label{fig:confidence_residual_llpic}}
    \caption{Coverage (left) and conditional confidence difference (right) as obtained on the \ac{LLPIC} test data, when training the \ac{ML} classifier and calibrating \ac{CP} on the \ac{MDC} data. The shaded grey area in the left plot represents the 99\% binomial confidence band corresponding to the minimum per-label effective calibration size $\min_y \hat{n}_y$. The dashed black vertical line in the right figure indicates the \ac{FAR} threshold of 1 per year.} 
    \label{fig:llpic}
\end{figure}

We then apply both standard and weighted \ac{CP}, and present the coverage results in \cref{fig:coverage_llpic}. We observe that standard \ac{CP} leads to undercoverage. Using weighted \ac{CP} improves the coverage toward the diagonal, but some undercoverage outside the expected binomial confidence band remains, suggesting that the observed distribution shift is not purely covariate. 

The conditional confidence residual is plotted in \cref{fig:confidence_residual_llpic}, showing that the conditional confidence increases when applying weighted \ac{CP} compared to the standard \ac{CP}. This is comparable to the induced covariate shift test case when resampling to lower feature values explored in \cref{fig:confidence_residual}, suggesting that the \ac{LLPIC} data contains a greater proportion of events with lower feature values relative to the \ac{MDC} calibration data. 
Examining the raw feature distributions, we find that, overall, the features in the \ac{LLPIC} data tend toward lower values than the corresponding \ac{MDC} values; however, this trend is not uniform, and the direction and magnitude of the shift depend on the specific feature and class considered.
Similar to the induced shift test cases, we also observe the greatest differences in conditional confidence for events with \maxifar values around the \ac{FAR} threshold, while noise events with small \maxifar and signals with high \maxifar vary little. 

\begin{figure}[h!]
    \centering
    \subfigure[Standard \ac{CP}]{\includegraphics[width=0.45\textwidth]{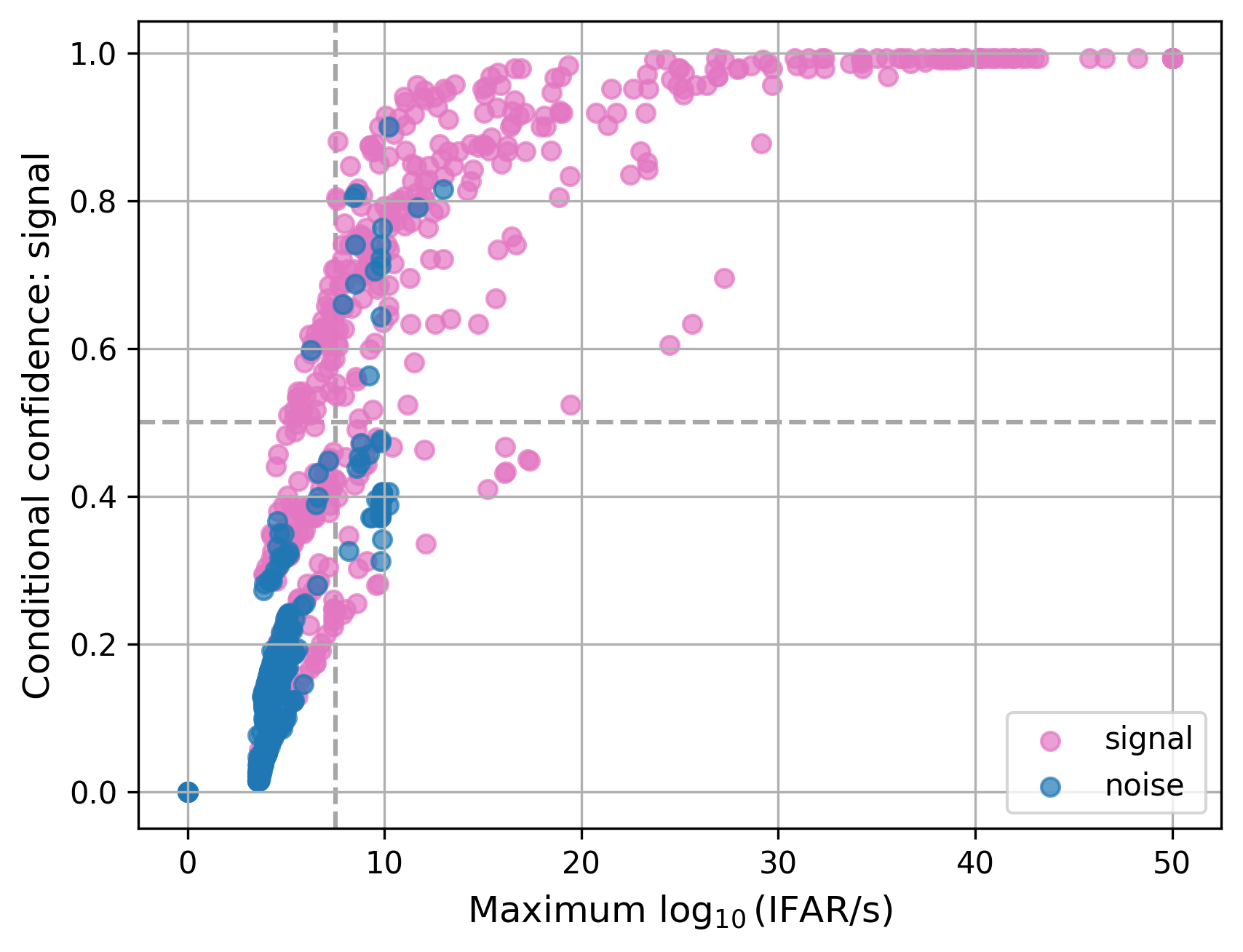}\label{fig:confidence_llpic_standard}}
    \subfigure[Marginal weighted \ac{CP}]{\includegraphics[width=0.45\textwidth]{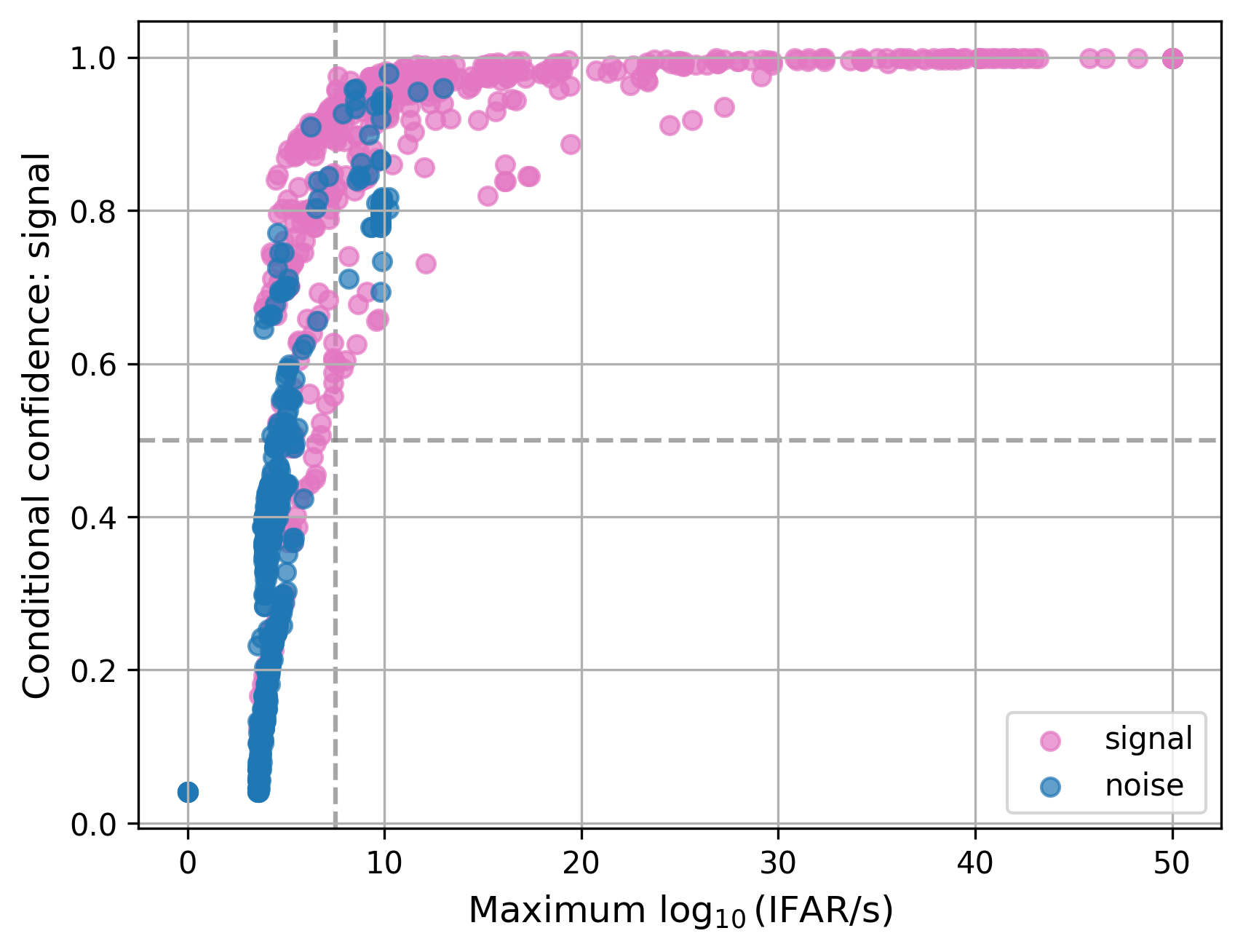} \label{fig:confidence_llpic_weighted}}
    \caption{Conditional confidence versus \maxifar on the LLPIC test data, as trained and calibrated on the MDC data, using standard Mondrian \ac{CP} (left) and weighted Mondrian \ac{CP} (right). The data are coloured by the true label.} 
    \label{fig:llpic_confidence}
\end{figure}

Finally, we investigate how the physical interpretations change when applying the weighted \ac{CP}. Defining a threshold of $0.5$ on the conditional confidence to distinguish between signal and noise predictions, we compare the resulting classifications against the \maxifar threshold corresponding to a \ac{FAR} of 1 per year, see \cref{fig:llpic_confidence}. The dataset contains a total of $534$ injected signals, of which $356$ are above the \maxifar threshold. 
Using standard \ac{CP}, we obtain $333$ predicted signals, corresponding to a \ac{TPR} of $61\%$ and a \ac{FPR} of $2\%$. 
However, weighted \ac{CP} gives $552$ signal predictions, increasing the \ac{TPR} to $84\%$ at the cost of a higher \ac{FPR} of $16\%$.

The regions of greatest interest for our \ac{GW} analysis are the quadrants where the conditional confidence and \maxifar disagree. 
In the upper left quadrants, events are classified as noise by \maxifar but as signals by our framework. Using standard \ac{CP} (see \cref{fig:confidence_llpic_standard}), there are $44$ such events, of which $43$ are true signals, while for the weighted \ac{CP} (see \cref{fig:confidence_llpic_weighted}), there are $141$ true signals among $196$ candidates. Thus, while the weighted \ac{CP} results in six times as many false positives (from $16$ to $106$), we also find $98$ new signals. 
In the bottom right quadrants, events with \maxifar above threshold but conditional confidence below threshold are located. We find $67$ such events for the standard \ac{CP}, including $31$ signals, while there are none for the weighted \ac{CP}. 

Overall, weighted \ac{CP} improves sensitivity to \ac{GW} signals at the cost of a higher false positive rate. By correcting for the covariate shift between the training and test distributions, the weighted \ac{CP} recovers many additional true signals that would otherwise be missed by standard \ac{CP}, reducing the false negative rate from $39\%$ to $16\%$. However, this comes at the expense of a sixfold increase in false positives, reflecting the reduced effective calibration sample size $\hat{n}_y$ incurred by the weighting procedure. The choice between standard and weighted \ac{CP} therefore represents a trade-off between sensitivity and purity that depends on the scientific goal: based on our empirical results, standard \ac{CP} is more conservative and better suited to constructing a high-purity catalogue, while weighted \ac{CP} is preferable when maximising the recovery of true signals is the priority.

\section{Conclusion} \label{sec:conclusion}
In this work, we have presented a framework for combining the outputs of multiple gravitational-wave search algorithms using Mondrian \ac{CP}, extended to account for distribution shift by reweighting calibration points according to the estimated likelihood ratio between the test and calibration feature distributions.
Using controlled distributional shifts in the simulated \ac{MDC} dataset, we demonstrated that standard \ac{CP} loses its coverage guarantee under covariate shift, with the degree of miscoverage scaling with the shift intensity. 
Weighted \ac{CP} successfully restores well-calibrated coverage across all shift intensities considered, confirming that the reweighting procedure can fully correct for induced covariate shifts of the type explored here.
We estimate the marginal likelihood ratio from unlabelled test data, which our empirical investigation shows to be the most robust choice, as class-conditional approximations rely on proxy labels whose accuracy cannot be guaranteed.

Applying the same framework to the unknown shift between two mock datasets, the \ac{MDC} and \ac{LLPIC}, we find that weighted \ac{CP} improves coverage toward the diagonal, though some residual undercoverage remains. This suggests the distributional difference between the two datasets is not purely a covariate shift, and that violations of the assumption that the class-conditional and marginal ratios are approximately equal may be present. The reweighting of the \ac{CP} quantile increases the conditional confidence of events, particularly for those near the \ac{FAR} threshold where classification is most uncertain.

In terms of physical interpretation, weighted \ac{CP} reduces the false negative rate, recovering signals that standard \ac{CP} would miss, at the cost of a higher false positive rate. This reflects an inherent sensitivity-purity trade-off: based on our empirical results, weighted \ac{CP} is well-suited to applications where completeness is paramount, such as population studies sensitive to selection effects, while the more conservative standard \ac{CP} remains preferable for constructing high-purity catalogues. 
When applied in practice to real detector data with unknown labels, the choice of method carries direct scientific consequences, as both missed signals and false detections can bias astrophysical conclusions.
Future work could explore tighter control of this trade-off, for example, through adaptive conditional confidence thresholds or improved likelihood ratio estimation for more complex distributional shifts.

\section*{Acknowledgements}
We want to thank Michael Coughlin, Deep Chatterjee, Tito Dal Canton, Reed Essick, Shaon Ghosh, Sushant Sharma-Chaudhary, Max Trevor, and Andrew Toivonen for the development of the \ac{MDC} results used in this work. 
We use data from \citet{LLPIC:2026prep} and thank the authors for preparing this dataset, and Sushant Sharma-Chaudhary and Thomas Sainrat for their help and advice in utilising it. 
Implementation of our \ac{ML} model was done using \sklearn \citep{Buitinck:2013fcp}, while we utilise \texttt{NumPy} \citep{harris_2020}, \texttt{Pandas} \citep{reback2020pandas}, and \texttt{Matplotlib} \citep{Hunter:2007} for data handling and visualisation. 
This material is based upon work supported by NSF’s LIGO Laboratory, which is a major facility fully funded by the National Science Foundation. 
The authors are grateful for computational resources provided by the LIGO Laboratory and supported by National Science Foundation Grants PHY-0757058 and PHY-0823459.
This work is supported by the Science and Technology Facilities Council (STFC) grant UKRI2488.

\bibliography{bibliography}

\end{document}